%Paper: hep-th/9408169
%From: Konstadinos Sfetsos <sfetsos@fys.ruu.nl>
%Date: Tue, 30 Aug 1994 17:54:38 +0100
%Date (revised): Tue, 30 Aug 1994 18:18:15 +0100
%Date (revised): Sun, 4 Sep 1994 14:28:50 +0100
%Date (revised): Wed, 7 Dec 1994 16:26:14 +0200

%%%%%%%%%%%%%%%%%%%%%%%%%%%%%%%%%%%%%%%%%%%%%%%%%%%%%%%%%%%%%%%%

\input harvmac

%\draftmode
\noblackbox
\baselineskip 20pt plus 2pt minus 2pt

\overfullrule=0pt

%%%%%%%%%%%%%%%%%%%%%%%%%%%%%%%%%%%%%%%%%%%%%%%%%%%%%%%%%%%%%%%%

% definitions

\def\bs{\bigskip}

\def\hb{\hfill\break}
\def\qq{\qquad}
\def\bl{\bigl}
\def\br{\bigr}

\def\IR{\relax{\rm I\kern-.18em R}}

%%%%%%%%%%%%%%%%%%%%%%%%%%%%%%%%%%%%%%%%%%%%%%%%%%%%%%%%%%%%%%

\def\r{\rho}
\def\a{\alpha}

\def\G{\Gamma}
\def\d{\delta}

\def\e{\epsilon}

\def\p{\pi}

\def\th{\theta}

\def\m{\mu}
\def\n{\nu}

\def\Om{\Omega}
\def\l{\lambda}
\def\L{\Lambda}
\def\s{\sigma}

\def\IR{\relax{\rm I\kern-.18em R}}

\def \bd {\bar \del}

\def \ha {{1\over 2}}

\def \ov {\over}

\def\sign{{\rm sign }}

%%%%%%%%%%%%%%%%%%%%%%%%%%%%%%%%%%%%%%%%%%%%%%%%%%%%%%%%%%%%%%%%
\def\fpi { f_{,i} }
\def\Fpi { F_{,i} }

\def\Fpu { F_{,u} }
\def\Fpv { F_{,v} }
\def\Apu { A_{,u} }
\def\Apv { A_{,v} }
\def\gpu { g_{,u} }
\def\gpv { g_{,v} }

\def\SW {Schwarzschild}
\def\RN {Reissner-Nordstr\"om}
\def\CS {Christoffel}
\def\MI {Minkowski}

%%%%%%%%%%%%%%%%%%%%%%%%%
%references

\lref\CALLAN{ C.G. Callan, D. Friedan, E.J. Martinec and M. Perry,
Nucl. Phys. {\bf B262} (1985) 593.}

\lref\WIT{E. Witten, Phys. Rev. {\bf D44} (1991) 314.}

\lref\Brinkman{H.W. Brinkmann, Math. Ann. {\bf 94} (1925) 119.}

\lref\GM{G. Gibbons, Nucl. Phys. {\bf B207} (1982) 337.\hb
G. Gibbons and K. Maeda, Nucl. Phys. {\bf B298} (1988) 741.}

\lref\KOULU{C. Kounnas and D. L\"ust, Phys. Lett. {\bf 289B} (1992) 56.}

\lref\DHooft{T. Dray and G. 't Hooft, Nucl. Phys. {\bf B253} (1985)
173.}
\lref\MyPe{ R.C. Myers and M.J. Perry, Ann. of Phys. {\bf 172} (1986)
304.}
\lref\GiMa{G. Gibbons and K. Maeda, Nucl. Phys. {\bf B298} (1988) 741.}
\lref\KLOPV{R. Kallosh, A. Linde, T. Ort\'\i n, A. Peet and A. Van Proeyen,
Phys. Rev. {\bf D46} (1992) 5278.}

%%%%%%%%%%%%%%%%%%%%%%%%%%%%%%%%%%%%%%%%%%%%%%%%%%%%%%%%%%%%%%%%%%%%%

%Send draft to: ykiem@puhep1.princeton.edu

%%%%%%%%%%%%%%%%%%%%%%%%%%%%%%%%%%%%%%%%%%%%%%%%%%%%%%%%%%%%%%%%%%%

\lref\lousan{ C.O. Loust\'o and N. S\'anchez, Int. J. Mod. Phys. {\bf A5}
(1990) 915; Nucl. Phys. {\bf B355} (1991) 231.}

\lref\lousanII{ C.O. Loust\'o and N. S\'anchez, Nucl. Phys. {\bf B383} (1992)
377.}

\lref\DHooftII{ T. Dray and G. 't Hooft, Commun. Math. Phys. {\bf 99}
(1985) 613.}

\lref\DHooftIII{T. Dray and G. 't Hooft,
Class. Quant. Grav. {\bf 3} (1986) 825.}

\lref\VV{H. Verlinde and E. Verlinde, Nucl. Phys. {\bf B371} (1992)
246.}

\lref\aisexl{P.C. Aichelburg and R.U. Sexl, Gen. Rel. Grav. {\bf 2}
(1971) 303.}

\lref\banados{M. Ba\~nados, C. Teitelboim, J. Zanelli, Phys. Rev. Lett. {\bf
69}
(1992) 1849; M. Ba\~nados, M. Henneaux, C. Teitelboim, J. Zanelli,
Phys. Rev. {\bf D48} (1993) 1506.}
\lref\hwka{G. Horowitz and D. Welch, Phys. Rev. Lett. {\bf 71} (1993) 328;
N. Kaloper, Phys. Rev. {\bf D48} (1993) 2598.}

\lref\cagan{C.G. Callan and Z. Gan, Nucl. Phys. {\bf B272} (1986) 647.}

\lref\hooft{ G. 't Hooft, Nucl. Phys. {\bf B335} (1990) 138.}
\lref\hoo{G. 't Hooft, Phys. Lett. {\bf B198} (1987) 61;
Nucl. Phys. {\bf B304} (1988) 867.}

\lref\tasepr{ I.S. Grandshteyn and I.M. Ryznik, {\it Tables of integrals,
Series and Products}, Academic Press (1980). }

\lref\HOTA{ M. Hotta and M. Tanaka, Class. Quantum Grav. {\bf 10} (1993) 307.}

\lref\otth{ V. Ferrari and P. Pendenza, Gen. Rel. Grav. {\bf 22} (1990)
1105.\hb
H. Balasin and H. Nachbagauer, ``The Ultrarelativistic Kerr-Geometry and
its Energy-Momentum Tensor'', gr-qc/9405053. \hb
K. Hayashi and T. Samura, Phys. Rev. {\bf D50} (1994) 3666.}
%``Gravitational shock waves for Schwarzschild and Kerr black holes'',
%KOBE-FHD-93-03A, gr-gc/9404027.}
%``Planckian scattering of massive particles and gravitational
%shock waves'', KOBE-FHD-93-05A, hep-th/9405013.}

\lref\KSHM{ D. Kramer, H. Stephani, E. Herlt and M. MacCallum,
{\it Exact Solutions of Einstein's Field Equations}, Cambridge (1980).}

%\lref\lou{ C.O. Lousto, ``The emergence of an effective two-dimensional
%Quantum Description from the study of Critical Phenomena in Black Holes'',
%gr-gc/9405048.}

\lref\lathos{ C.O. Loust\'o and N. S\'anchez, Phys. Lett. {\bf B220} (1989)
55.}

\lref\FMS{D. Friedan, E. Martinec and S. Shenker, Nucl. Phys. {\bf B271} (1986)
93.}

\lref\BAIT{ M. Bander and C. Itzykson, Rev. of Mod. Phys. {\bf 38} (1966) 330;
Rev. of Mod. Phys. {\bf 38} (1966) 346.}

\lref\rpen{R. Penrose, in General Relativity: papers in honour of J.L. Synge,
ed. L. O'Raifeartaigh (Clarendon, Oxford, 1972) 101.}

\lref\AGV{L.N. Lipatov, Nucl. Phys. {\bf B365} (1991) 614.\hb
R. Jackiw, D. Kabat and M. Ortiz,
Phys. Lett. {\bf B277} (1992) 148.\hb
D. Kabat and M. Ortiz, Nucl. Phys. {\bf B388} (1992) 570.\hb
D. Amati, M. Giafaloni and G. Veneziano,
Nucl. Phys. {\bf B403} (1993) 707.\hb
M. Fabbrichesi, R. Pettorino, G. Veneziano and G.A. Vilkovisky,
Nucl. Phys. {\bf B419} (1994) 147.}

%F.M. Paiva, M.J. Reboukas and M.A.H. MacCallum,
%Class. Quant. Grav. {\bd 10} (1993) 1165;

%%%%%%%%%%%%%%%%%%%%%%%%%%%%%%%%%%%%%%%%%%%%%%%%%%%%%%%%%%%%%%%%
% begin paper

\hfill {THU-94/13}
\vskip -.3 true cm
\rightline{August 1994}
\vskip -.3 true cm
\rightline {hep-th/9408169}

\bs\bs\bs

\centerline  {\bf ON GRAVITATIONAL SHOCK WAVES IN CURVED SPACETIMES }
%\centerline  { \bf  }

\vskip 1.00 true cm

\centerline  {  {\bf Konstadinos Sfetsos}{\footnote{$^*$}
 {e-mail address: sfetsos@fys.ruu.nl }}                                     }

\bigskip

\centerline {Institute for Theoretical Physics }
\centerline {Utrecht University}
\centerline {Princetonplein 5, TA 3508}
\centerline{ The Netherlands }

%%%%%%%%%%%%%%%%%%%%%%%%%%%%%%%%%%%%%%%%%%%%%%%%%%%%%%%%%%%%%%%%%%%%

\vskip 1.30 true cm

\centerline{ABSTRACT}

Some years ago Dray and 't Hooft found the necessary and sufficient conditions
to introduce a gravitational shock wave in a particular class of vacuum
solutions to Einstein's equations. We extend this work to cover cases where
non-vanishing matter fields and cosmological constant are present.
The sources of gravitational waves are massless particles moving along a null
surface such as a horizon in the case of black holes.
After we discuss the general case we give many explicit examples.
Among them are the $d$-dimensional charged
black hole (that includes the 4-dimensional Reissner-Nordstr\"om and the
$d$-dimensional Schwarzschild solution as subcases),
the 4-dimensional De-Sitter and Anti-De-Sitter spaces
(and the Schwarzschild-De-Sitter black hole), the 3-dimensional
Anti-De-Sitter black hole, as well as
backgrounds with a covariantly constant null Killing vector.
We also address the analogous problem for string inspired
gravitational solutions and give a few examples.

\vskip .3 true cm

\vfill\eject

%%%%%%%%%%%%%%%%%%%%%%%%%%%%%%%%%%%%%%%%%%%%%%%%%%%%%%%%%%%%%%%%%%%%

\newsec{ Introduction }

In an $S$-matrix approach to black hole physics \hooft\
one has to take into
account the interactions between Hawking particles emitted from the black hole.
However in order to do that one would like also to know
how the presence of particles around a black hole affects the black hole itself
(the black hole reacts back).
It has been convincingly argued \hooft\ that the gravitational
interaction between particles at Planckian energies (see \hoo\lousan\VV\AGV)
dominates any other
type of interaction and therefore one needs to know what gravitational effects
the particles have on the original black hole geometry.

For the case of a massless particle moving along the horizon of a \SW\ black
hole the result of this elementary classical black hole back reaction
has been found by Dray and 't Hooft \DHooft.
What they found generalized the gravitational shock wave due to
a massless particle moving in flat \MI\ space \aisexl.
In fact \DHooft\ contains
the necessary and sufficient conditions for being able to introduce a
gravitational shock wave via a coordinate shift
in more general {\it vacuum}
solutions to Einstein's equations.\foot{ The solution of
\DHooft\ can be formulated as two \SW\ black holes of {\it equal masses}
glued together at the horizon. For a spherical shell of massless
matter moving along
$u=u_0\neq 0$ the solution \DHooftII\ represents two \SW\
black holes of {\it unequal masses} glued together at $u=u_0$.
Various gravitational shock waves
obtained by infinitely boosting \aisexl\DHooft\
known solutions have been found
(see for instance \lousan\lousanII\HOTA\otth).}

The purpose of this paper is to extend the results of \DHooft\ to the case when
matter fields and even a non-vanishing cosmological constant are present since,
many interesting gravitational solutions belong in this category.
The organization of this paper and some of the important results are:
In section 2 we address the problem of how a quite general class of
solutions to Einstein's
equations in the {\it presence of matter} and {\it cosmological constant}
(in $d$ dimensions) back reacts to a
massless particle moving along a null surface.
We also examine the geodesics of particles moving in these geometries
and show that discontinuity and refraction effects take place as they try
to cross the null surface.
%In particular the latter effect has not been
%discussed before in the literature in curved spacetime.
Our treatment, as the one in \DHooft, is non-perturbative but exact.
In section 3 we consider the case of the $d$-dimensional \RN\ charged
black hole \MyPe.
After we give the general result we concentrate to the 4-dimensional case
which physically is the most interesting one.
We find that as we approach the extremal case of equal mass and charge
the effect the massless particle has become gigantic in magnitude.
In section 4 we consider firstly the De-Sitter and Anti-De-Sitter constant
curvature spaces in four dimensions. In the former case we find that the
discontinuity in geodesics we have already mentioned depends in a rather
unexpected way on the angular distance from the position of the massless
particle. We also consider the case of the \SW-De-Sitter black hole which
interpolates between the De-Sitter space and the \SW\ black hole.
We close this section with the 3-dimensional Anti-De-Sitter space which,
after a discrete identification, can be interpreted as a 3-dimensional black
hole \banados.
In section 5 we consider the analogous problem for string inspired
gravitational solutions. In particular for a 4-dimensional electrically and
magnetically charged dilatonic black hole \GiMa\KLOPV\
and the background corresponding
to the conformal field theory $SL(2,\IR)/\IR \otimes \IR^2$.
% i.e. the direct product of the 2-dimensional black hole \WIT\ with
%2 extra flat dimensions.
We end the paper with concluding remarks and discussion in section 6.
Mainly in order not to interrupt the flow of the paper with too many
mathematical details we have written Appendices A, C and D. In Appendix B
we consider the case where the background geometry has a covariantly
constant null Killing vector \Brinkman\
(these geometries do not belong to the class
considered already in section 2).

\newsec{  General results }

Let us consider the $d$-dimensional spacetime described by the
metric
\eqn\metrd{ ds^2= 2\ A(u,v)\ du dv\ +\ g(u,v)\ h_{ij} (x)\ dx^i dx^j\ ,}
with $ (i,j=1,2,\dots , d-2) $.
Let us also
assume that there exist some matter fields with non-vanishing components
of the energy momentum tensor given by
\eqn\enemom{ T= 2\ T_{uv}(u,v,x)\ dudv + T_{uu}(u,v,x)\ du^2
+ T_{vv}(u,v,x)\ dv^2
 + T_{ij}(u,v,x)\ dx^idx^j\ .}
Notice that this form of energy momentum tensor is consistent with the Ricci
tensor for \metrd\ as given by (A.2) for $f=0$.

We consider a massless particle located at $u=0$ and
moving with the speed of light in the $v$-direction and
we want to find out what its effect is on the geometry described by \metrd.
Similarly to \DHooft\ our ansatz will be that for $u<0$ the spacetime is
described by \metrd\
and for $u>0$ by \metrd\ but with $v$ shifted as $v\to v+f(x)$, where $f(x)$
is a function to be determined.
Therefore the resulting spacetime metric and energy momentum tensor are
\eqn\metrgl{ ds^2=  2\ A(u,v+\Theta f )\ du (dv + \Theta \fpi dx^i)
\ +\ g(u,v+\Theta f)\ h_{ij} (x)\ dx^i dx^j\ ,}
where $\Theta=\Theta(u)$ is the Heaviside's step function
and\foot{For spherical sourceless
shock waves in \MI\ space obtained with a different ansatz than
\metrgl\ see \rpen.}
\eqn\ener{\eqalign{T =  & 2\ T_{uv}(u,v+\Theta f,x)\ du(dv +\Theta \fpi dx^i)
+ T_{uu}(u,v+\Theta f,x) du^2 \cr
&+ T_{vv}(u,v+\Theta f,x)\ (dv +\Theta \fpi dx^i)^2 +
T_{ij}(u,v+\Theta f,x)\ dx^idx^j\ .\cr }}
In order to compute various tensors it would be easier to transform to the
new coordinates
\eqn\transf{ \hat u= u\ ,\qq \hat v = v + f(x) \Theta (u) \ ,
\qq \hat x^i = x^i\ ,}
in which the metric \metrgl\ and the energy momentum tensor take the form
\eqn\metrhat{\eqalign{& ds^2 = 2\ \hat A\ d\hat u d\hat v\ + \
\hat F\ d\hat u^2\ +\
\hat g\ \hat h_{ij}\ d\hat x^i d\hat x^j \cr
& \hat F= F(\hat u, \hat v, \hat x)= -2\ \hat A\ \hat f\ \hat \d\ ,\cr } }
and
\eqn\enhat{ T= 2\ (\hat T_{\hat u\hat v} - \hat T_{\hat v \hat v} \hat
f \hat \d ) \ d\hat u d\hat v +
(\hat T_{\hat u \hat u} + \hat T_{\hat v \hat v} \hat f^2 \hat \d^2
- 2 \hat T_{\hat u\hat v} \hat f \hat \d)\ d \hat u^2 + \hat T_{\hat v
\hat v}\ d \hat v^2
+ \hat T_{ij}\ d\hat x^i d \hat x^j\  ,}
where the hats indicate that the corresponding quantities are evaluated
at $\hat u$, $\hat v$, $\hat x^i$
 and where $\hat \d=\d (\hat u)$ is the $\d$-function.

The various metric and field components must satisfy
Einstein's equations in the presence of matter, which in
$d$-dimensions read (throughout this paper $c=\hbar=G=1$)
\eqn\Einst{ R_{\m\n} - \ha g_{\m\n} R = -8\pi \ T_{\m\n} \
\Rightarrow \  R_{\m\n} = - 8\pi \ (T_{\m\n}
- {1\ov d-2}\ g_{\m\n}\ T_{\l}{}^{\l}) \equiv -8\pi \ \tilde T_{\m\n}\ .}
Obviously $\tilde T= \tilde T_{\m\n} dx^{\m} dx^{\n}$ is also
of the form \enemom.
We would like to find solutions of \Einst\ with metric tensor of
the form \metrhat\ and an energy momentum tensor given by \enhat\ plus the
contribution of the energy momentum tensor for a massless particle
located at the origin of the transverse $x$-space and at $u=0$ and moving with
the speed of light in the $v$-direction
\eqn\enepart{ T^{\rm p} = T^p_{uu}\ du^2 = \hat T^{\rm p}_{\hat u \hat u}\
d\hat u^2=
-4\ p\ \hat A^2\ \hat \d^{(d-2)}(\hat x)\ \hat \d(\hat u)\ d\hat u^2\ ,}
where $p$ is the momentum of the particle.
All the relevant tensor components appearing in \Einst\ are given in
Appendix A. To simplify the notation we will also drop the hats over the
symbols keeping in mind however the transformation \transf.
Assuming that the parts of the equations \Einst\ that do not involve
the function $f$ are satisfied, one finds by
examining the linear in $f\d$ terms that at $u=0$
the additional conditions
\eqn\condit{\eqalign{ & \gpv = \Apv =  T_{vv} = 0 \cr
& \triangle_{h_{ij}} f \ - \ {d-2\ov 2}\ {g_{,uv}\ov A}\ f \ =\
32\pi \ p\ g\ A\ \d^{(d-2)}(x)\ ,
\cr }}
must also be satisfied, where the Laplacian is defined as
$\triangle_{h_{ij}}= 1/\sqrt{h} \del_i \sqrt{h} h^{ij}\del_j$.
In order to cast the differential equation in \condit\ into the given form
we used the fact that at $u=0$
\eqn\auv{
{A_{,uv}\ov A} \ =\ -{d-2\ov 2}\ {g_{,uv}\ov g}\ + \ 8\pi \ \tilde T_{uv}\ .}
This equality follows from the $(\m,\n)=(u,v)$ components of \Einst\ computed
at $u=0$ and for $f=0$.
Notice also that the differential equation in \condit\ does not explicitly
depend on the
components of the energy momentum tensor of the matter fields. Its dependence
on these fields is only implicit through the functions $A$, $g$ that
are determined from the $f$-independent part of Einstein's equations.

Next we examine the $f^2 \d^2$ type terms. This is important because
such terms should also vanish (in a distribution sense), otherwise our
considerations are perturbative in powers of $f$.
Using the first line in \condit\ it is easy to see
that the coefficient of $f^2 \d^2$ in the $R_{uu}$ component of the
Ricci tensor in (A.2) has terms of order $O(u)$ and $O(u^2)$ (in all
of our examples such terms are of order $O(u^2)$ since there is
functional dependence only on the product $uv$).
Remembering that we are
really considering all the quantities involving $\d$-functions as
distributions to be integrated over smooth functions we find that in
this case such an integral vanishes. Moreover because $ T_{vv}=0$ at
$u=0$ we have that $ T_{vv}= O(u)$ (at least).
Therefore we can take the terms in \Einst\
involving $f^2\d^2$ to zero.
A last crucial remark is that because we have assumed that $f$ is a
function of the $x^i$'s only, the potential $v$ dependence in \condit\ should
drop out for a consistent solution to exist. Mathematically that
implies that the coefficient of the order $O(u)$ term in the
expansion of $g(u,v)$ in powers of $u$ should be a linear function of
$v$. In fact this is the case in all the examples we explicitly work out.

For the 4-dimensional case in the absence of matter fields the condition
\condit\ was given in \DHooft. It is remarkable that \condit\
has essentially the same form
in the vacuum and in the presence of matter except that
in the latter case the additional condition $ T_{vv} =0$ at $u=0$ should
be imposed as well. Let us also emphasize that the differential equation in
\condit\ is nothing but the Green function equation (of course with
$\d^{(d-2)}(x)$ replaced by $\d^{(d-2)}(x-x')$). Thus
instead of a point massless particle we could easily find the result if we
had an extended source with density $\r(x)$,
i.e. $f(x)=\int \r(x) f(x,x') d x'$.
In this way we can consider spherical and planar shells of matter as
in \DHooftII\DHooftIII\ or charged particles, cosmic strings,
monopoles etc. (for the flat space case see \lousan).

The following remark is now in order. According to the work of \VV\ in the
4-dimensional case in the equation for $f(x)$
in \condit\ half (in our normalization) the 2-dimensional curvature
$R^{(2)}$ constructed out of the metric $h_{ij}$ should be in place of the
factor ${g_{,uv}\ov A}$. In our case using \Einst\ for $(\m,\n)=(i,j)$, the
fact that $\gpv =0$ at $u=0$ and (A.2) we compute that at $u=0$
\eqn\RR{ {d-2\ov 2}\ {g_{,uv}\ov A} = \ha R^{(d-2)} + 4\pi \ \tilde
T_{ij} h^{ij}\ .}
Thus we obtain the result of \VV, and in fact generalized in any
number of spacetime dimensions, if $\tilde T_{ij}h^{ij} =0$ or, after using
\Einst\enemom, if $T_{uv}=0$.
The stronger conditions
$T_{ij}=0$ and $T_{\l}{}^{\l}=0$, which imply $\tilde T_{ij}=0$ and in our case
$T_{uv}=0$ were imposed in \VV. However, in many interesting cases, such as the
\RN\ charged black hole, the component $T_{uv}$ is non-vanishing
as we shall see.

Although the explicit form of the metric, once the function $f(x)$ is obtained
by solving \condit, is given by \metrhat\ it would be helpful to examine the
geodesic equations in order to obtain a clearer understanding of how the
original geometry \metrd\ is affected by the presence of a massless particle
moving in the $v$-direction at $u=0$.
It is shown in Appendix C that as the geodesic trajectory
crosses the null surface $u=0$ there is a shift in its $v$-component
given by
\eqn\shifftt{ \Delta v = f(x) \ ,}
as well as a refraction effect in the transverse $x$-plane expressed by
the `refraction function'
\eqn\refra{ R(x) \equiv
{dx^i\ov du}\big |_{u=0^-}  -  {dx^i\ov du}\big |_{u=0^+}  =
{A \ov g}\big |_{u=0}\ f_{,j} h^{ji} \ ,}
that essentially measures how much the angle that the trajectory forms
with the
$u=0$ surface changes as we cross this surface.
In other words a generic trajectory when it crosses $u=0$ suffers the
discontinuity \shifftt\ in its $v$-component, with the $u$ and $x^i$ components
being continuous at that surface, and
moreover its $x^i$-components change
direction along $u$ according to \refra.
Although in both phenomena $f(x)$ determines the functional dependence
there is however a qualitative difference.  In the case of \shifftt\ the
discontinuity equals $f(x)$ whereas in \refra\ the directional derivatives
of $f(x)$ play the important role. Therefore there might be points $x^i$
where there is a discontinuity in the trajectory but no refraction effect
and vice versa. In fact this will be the case in some of our examples
considered in the following sections.

\newsec { The $d$-dimensional Reissner-Nordstr\"om charged black hole}

With this section we shall start applying the general formalism that we
developed in the previous one by considering the
case of a massless particle moving in the outer horizon of a
\RN\ charged black hole in $d$ dimensions.
Since the latter is a spherically symmetric,
asymptotically flat solution of \Einst,
our results will be a natural extension of those obtained by
Dray and 't Hooft for the case of the 4-dimensional \SW\ black
hole. The metric for the \RN\ charged black hole solution in
$d$ dimensions is \MyPe
\eqn\rei{\eqalign{& ds^2 = - \l (r)\ dt^2 \ + \ \l^{-1} (r) \ dr^2 \
+ \ r^2 d\Omega^2_{(d-2)}  \cr
& \l(r) = 1- {2C\ov r^{d-3} } + {E^2\ov r^{2(d-3)}} \ ,\cr } }
where $C$, $E$ are constants depending on the dimension of spacetime
$d$ and related to the mass and the
charge of the black hole (see \MyPe).
The energy momentum tensor of the electromagnetic field is given by
\eqn\electro{ T_{\m\n}={d\ov 8\pi  (d-2)} \bl( F_{\m\l} F_{\n}{}^{\l}
-{1\ov d} g_{\m\n} F^2 \br)\ ,
\qq F_{tr}=\sqrt{\ha (d-2)(d-3)}\ {E\ov  r^{d-2}}\ ,}
where $\m,\n=t,r,i$ and $F_{tr}$ denotes the only non-vanishing component of
the electromagnetic field 2-form.
It is possible to bring \rei\ into the form \metrd\
by means of the following transformation
\eqn\traa{ u= e^{t/\a} F(r)\ ,\qq v = e^{-t/\a} F(r)\ , \qq
\a =  {2 r_+ \ov d-3} \bl( 1- ({r_- \ov r_+})^{d-3}\br)^{-1} \ ,}
where $r^{d-3}_{\pm}= C \pm \sqrt{C^2 -E^2}$, with $r_+ > r_-$,
 are the outer and inner horizons where the function
$\l (r)$ in \rei\ vanishes.\foot{Then
since the only non-vanishing components of
the electromagnetic field would be $F_{uv}$, the corresponding energy momentum
tensor \electro\ would be of the form \enemom\ with $T_{uu}=T_{vv}=0$.
Nevertheless, its precise form
will not be needed for our purposes as we have already explained.}
It turns
out that
\eqn\Fun{ F(r)= \exp{ \bl( {1\ov \a} \int dr \l^{-1}(r) \br ) }
= \kappa (r-r_+)^{\ha} + \dots\,}
where $\kappa$ is a constant and the dots stand for higher order terms in the
$(r-r_+)$ expansion.\foot{ The explicit expression for $\kappa$ is quite
straightforward to write down in any number of dimensions. However the result
is complicated and not very enlightening. For $d=4$ the expression is simpler
\eqn\kaaa{\kappa= (r_+-r_-)^{-\ha {r_-^2\ov r_+^2}} e^{r_+ - r_-\ov 2
r_+}\ .} }
We also find that
\eqn\ag{ A(u,v)= \ha \a^2 \l(r) \exp{\bl( -{2\over \a} \int dr
\l^{-1}(r)\br) }\ , \qq g(u,v) = r^2 \ , }
and that at $u=0$
the conditions $\gpv=\Apv=0$ of \condit\ are indeed satisfied.
Using the values of the following functions at $u=0$
\eqn\aggg{ A= {2 r_+ \ov d-3} \kappa^{-2}
 \bl( 1- ({r_-\ov r_+})^{d-3} \br)^{-1}\ ,
\qq g= r_+^2 \ ,\qq g_{,uv}= 2 r_+ \kappa^{-2} \ ,}
we find from \condit\ that the equation the shift function satisfies is
\eqn\condd{  \triangle_{(d-2)} f \ - \ a(d,r_+,r_-)\ f \ =
 \ 2\pi \ b(d,r_+,r_-)\ \d^{(d-2)}(x) \ ,}
where $\triangle_{(d-2)}$ is the Laplacian on the unit $(d-2)$-sphere
and $a(d,R_+,r_-)$, $b(d,r_+,r_-)$ are constants defined as
\eqn\connn{\eqalign{
& a(d,r_+,r_-)\equiv \ha(d-2)(d-3) \bl( 1- ({r_- \ov r_+})^{d-3}\br) \cr
& b(d,r_+,r_-) \equiv 32p r_+^3 (d-3)^{-1} \kappa^{-2}
\bl( 1- ({r_- \ov  r_+})^{d-3}\br)^{-1} \ .\cr } }
Physically the most interesting is the 4-dimensional case.
Then \condd\ takes the form of (D.1) with
\eqn\coock{c=1-{r_-\ov r_+}\ \ ( 0< c \leq 1 )\ ,\qq
k= 32  p r_+^4 (r_+ - r_- )^{-{r_+^2- r_-^2\ov r_+^2}}
e^{-{r_+ - r_- \ov r_+ }} \ .}
Obviously if the black hole is uncharged (\SW),
i.e. $r_-=0$, we recover the result of
Dray and 't Hooft \DHooft.
%\foot{ The charged dependent constant $c$ in
%\coock\ gives rise to an indirect electromagnetic interaction
%between the uncharged massless particle at $r=r_+$ and the charged \RN\
%black hole. This pheneomenon is quite general, since the constant $c$ depends
%on $A(u,v)$, $g(u,v)$ that encode parameters of the original background
%geometry.}
The solution of the equation is given in (D.2).\foot{
The solution of \condd\ for the higher than $d=4$ dimensional cases
can easily be given in terms of generalized spherical functions
(see eqn. (D.12) in Appendix D).}
%these are representations functions of $SO(d-1)$) \BAIT.}
For ${1\ov 4} \leq c \leq 1$ (and in fact for all $c \geq {1\ov 4}$)
an integral representation (proportional to a hypergeometric series)
of the solution can be found using
\eqn\repres{ \int_0^{\infty} ds\ e^{-(l+\ha) s} \cos (\sqrt{c-1/4}\ s) =
{l+\ha\ov l(l+1) +c}\ , }
and the generating function for Legendre polynomials
\eqn\repp{ \sum_{l=0}^{\infty} t^l P_l(\cos \th) =
(1-2 \cos \th\ t + t^2 )^{-1/2} \ .}
We find (consult \tasepr)
\eqn\fff{\eqalign{ f(\th;c) &=
 -{k \ov \sqrt{2}} \int_0^{\infty} ds \cos(\sqrt{c-1/4}\ s)\
{1\ov \sqrt{\cosh s - \cos \th }} \cr
&= -{k \pi \ov 2 \cosh(\sqrt{c-1/4}\ \pi)}\
F(\ha -i\sqrt{c-1/4},\ \ha +i\sqrt{c-1/4},\ 1,\ \cos^2 {\th\ov 2})\ .\cr } }
%For $0<c\leq {1\ov 4}$ the $\cos(\sqrt{c-1/4}\ s)$ factor
%in \repres\fff\ should be replaced by $\cosh(\sqrt{1/4-c}\ s)$.
For $0<c\leq {1\ov 4}$ we should replace $\sqrt{c-1/4}$ by $i\sqrt{1/4-c}$
with parallel replacement of the corresponding trigonometric functions
by hyperbolic ones and vice versa.
Notice that the solution blows up at the point of the unit 2-sphere
where the particle was placed,
i.e. at $\th =0$. Moreover it is everywhere negative and for fixed $c$
it is a monotonically increasing function of
$\th \in [0,\pi]$ approaching a constant at $\th=\pi$.
For fixed $\th$ it also monotonically increases as a
function of $c \in (0,1 ]$. The `refraction function' is
\eqn\rth{ R(\th;c)= \bl({A\ov g}\br)_{u=0}\ \partial_{\th} f(\th;c)\ .}
As a function of $\th$ it monotonically decreases from plus infinity
at $\th=0$ to zero at $\th=\pi$.

It is useful to examine what happens in the region close to the
two poles of the sphere where the behavior is extreme.\foot{ In the
following $I_n(x)$ and $K_n(x)$ will denote, as usual,
the $n$th-order modified Bessel functions.
The relations: $K_0'=-K_1$, $I_0'=I_1$ and
the leading order behavior as $x<<1$: $I_0\simeq 1$, $I_1\simeq {x\ov 2}$,
$K_0 \simeq -\ln x$, $K_1 \simeq {1\ov x}$ will also prove useful.}
Close to the northern pole at $\th =0$ we have
\eqn\npole{ f(\th;c) \simeq \cases{ -k\ K_0(\sqrt{c}\ \th)\ , \quad
& as $ \th << 1$
\cr \noalign{\vskip 6pt}
k\ \ln(\sqrt{c}\ \th )\ ,\quad & as $\th <<< 1$\ ,\cr }}
whereas close to the southern pole at $\th=\pi$
\eqn\spole{ f(\th;c) \simeq \cases{ -{k\pi\ov 2\cosh(\sqrt{c-{1\ov 4}} \pi)}\
I_0(\sqrt{c}\ (\pi-\th))\ , \quad & as $ \pi-\th << 1$
\cr \noalign{\vskip 6pt}
 -{k\pi\ov 2\cosh(\sqrt{c-{1\ov 4}} \pi)}\ ,\quad & as $\pi-\th <<< 1$\ .\cr }}
For the `refraction function' the corresponding results are
\eqn\npolere{ R(\th;c) \simeq \cases{ k'\sqrt{c}\ K_1(\sqrt{c}\ \th)\ , \quad
& as $ \th << 1$
\cr \noalign{\vskip 6pt}
{k'\ov \th}\ ,\quad & as $\th <<< 1$\ ,\cr }}
and
\eqn\spolere{ R(\th;c) \simeq \cases{
{k'\pi \sqrt{c}\ov 2\cosh(\sqrt{c-{1\ov 4}} \pi)}\
I_1(\sqrt{c}\ (\pi-\th))\ , \quad & as $ \pi-\th << 1$
\cr \noalign{\vskip 6pt}
{k'\pi c\ov 4\cosh(\sqrt{c-{1\ov 4}} \pi)}\ (\pi-\th)\ ,
\quad & as $\pi-\th <<< 1$\ ,\cr }}
where $k'=(A/g)_{u=0} k$ is a positive constant.
As it was expected as we approach $\th=0$ both $f$ and
$R$ behave like the corresponding functions in the flat space
case \aisexl\DHooft\ (see also (B.8))
and both approach infinity. At the southern pole
both functions reach their minimum magnitudes
and in fact the refraction phenomenon
disappears even though a particle trajectory is still discontinuous since
$f(\pi;c)\neq 0$.

We end this section with a comment on the case of the extremally charged \RN\
black hole. It is obvious either from the series
representation (D.2) for $f(\th;c)$, or from the integral
one \fff\ that for $c=0$, or equivalently in the extremal limit
$r_-=r_+$, the solution is not well defined. In fact the choice
we have made for the constant $\a$ in \traa\ and everything that
follows are not valid if $r_-=r_+$. It can easily be shown that there
is no solution to our problem, or in other words no {\it single}
particle can move with the speed of light in the outer
horizon of an extremally charged
\RN\ black hole. Physically this should have been
expected because extremallity is the situation where the two horizons
coincide just before a naked singularity appears. From a more mathematical
point of view this happens because the function $F^2(r)$ is not
analytic at $r=r_+=r_-$, as it is in the non-extremal case (see \Fun).

%\newsec{ The cases of the $d=4$ and $d=3$ (Anti)-De-Sitter spaces}
\newsec{ Constant curvature spaces }

In this section we consider
the De-Sitter and Anti-De-Sitter spaces where there is a constant
curvature $R={2d/(d-2)} \L$ corresponding to a non-vanishing
cosmological constant $\L$. Even though a cosmological term was not
considered in \Einst\ its effect can be imitated by an energy momentum
tensor in \Einst\ that is proportional to $g_{\m\n}$,
i.e. $8\pi  \tilde T_{\m\n} = -2/(d-2) \L g_{\m\n}$.
We also analyze the \SW-De-Sitter black hole case.

\subsec{ The 4-dimensional cases }

The metric for the 4-dimensional De-Sitter space is
\eqn\desitter{ ds_{DS}^2 = -(1-{r^2\ov a^2})\ dt^2 + (1-{r^2\ov a^2})^{-1} dr^2
+r^2 (d\th^2 + \sin^2 \th\ d\phi^2) \ ,}
whereas for the corresponding Anti-De-Sitter one is
\eqn\adesitter{ ds_{ADS}^2 =
-({r^2\ov a^2}-1)\ dt^2 + ({r^2\ov a^2}-1)^{-1} dr^2
+r^2 (d\chi^2 + \sinh^2 \chi\ d\phi^2) \ .}
In both cases the constant $a$ is related to the non-vanishing
cosmological constant\foot{
Throughout this subsection upper (lower) signs in some quantities
correspond to the De-Sitter (Anti-De-Sitter) case.} as $\L = \pm 3/a^2 $.
To bring \desitter, \adesitter\
into the form \metrd\ we make the same coordinate
transformation as in \traa\ but with $\a= a$. Then it turns out that
\eqn\tutt{ F(r)=\bl(\pm{a-r\ov r+a}\br)^{\ha}\ ,\qq A(u,v)= \ha (r+a)^2\ ,
\qq g(u,v)=r^2\ .}
The conditions of \condit\ are indeed satisfied at $u=0$ ($r=a$) and
the differential equation $f$ satisfies is of the form (D.1) with
%\eqn\formm{ \triangle_{(2)} f + 2 f =- 2\pi k \d^{(2)}(x) \ ,\
%qq k=32 G p a^4 \ .}
\eqn\formm{  c=\mp \ 2\ ,\qq k= 32  p a^4 \ .}
Then according to (D.5) for $l=1$ the solution for the De-Sitter case is
%\eqn\desso{
%f_{DS}(\th) =  32  p a^4\ Q_1(\cos \th)\ \Theta ({\pi\ov 2}-\th)\
%, \qq Q_1(\cos \th)=\ha  \cos \th\ \ln (\cot^2 {\th\ov 2} ) - 1\ ,}
\eqn\desso{f_{DS}(\th) =  32  p a^4\
\bl(1- \ha  \cos \th\ \ln (\cot^2 {\th\ov 2} )  \br)\
\Theta ({\pi\ov 2}-\th)\ ,}
where contributions proportional to
$Y_1^m(\th,\phi)$, $m=-1,0,1$ that solve the homogeneous equation
have been omitted. As it also pointed out in the Appendix D the use of the
$\Theta$-function that essentially restricts the solution to the upper
hemisphere is necessary in order to have a solution that does not blow up at
the southern pole at $\th=\pi$, which would be unphysical since no particle
is placed there. Comparing with the black hole case we considered in the
previous section we find a notable difference. The solution goes to
minus infinity at $\th=0$ and then monotonically increases until it reaches
the value ($32pa^4$) at $\th={\pi \ov 2}$.
Therefore there is an angle ($\th_0 \simeq 33.52^0$)
where it becomes zero and then it passes from negative to positive values.
Thus contrary to intuition the magnitude of the shift has its minimum not
at $\th=\pi$ (as it was the case in the previous section) but at an
intermediate angle.
The corresponding `refraction function' is
\eqn\kost{ R_{DS}(\th)= 64 p a^4\
\bl( {\cos\th \ov \sin 2\th} + \ha \sin\th\ \ln (\cot^2 {\th\ov 2})\br)\
\Theta({\pi\ov 2}-\th) \ - \ 64 p a^4\ \d (\th-{\pi\ov 2})\ .}
The first term is a monotonically decreasing function of $\th$ and it varies
from plus infinity to zero as we
go from the northern pole to the equator. However exactly there the second
term gives an infinite contribution. This is in effect
a consequence of restricting the solution to the upper hemisphere and, in
some sense, can be thought of as a source term that displaces the minimum
magnitude of $f_{DS}(\th)$ from $\th={\pi\ov 2}$ to $\th=\th_0$.
Notice also that at $\th=\th_0$ there is no discontinuity for the geodesic
trajectories but there is still a refraction effect.

A natural question to ask is whether or not there is a distribution of massless
particles, instead of just a single one, for which the use of a
$\Theta$-function in \desso\ is unnecessary. Obviously there are many such
distributions (after all \desso\ is nothing but a Green function).
The simplest one is to consider two particles of the {\it same energy}
$p$ one in each
pole of the 2-sphere. Then the solution becomes exactly \desso\ with
no $\Theta$-function of course, and in fact it can be found by infinitely
boosting a \SW-de-Sitter black hole (see (4.9) below)
in the zero mass limit \HOTA.

For the Anti-De-Sitter case (D.8) for $l=1$ gives
%\eqn\adesso{ f_{ADS}(\chi) =
%32  p a^4\ Q_1(\cosh \chi)\ ,
%\qq Q_1(\cosh \chi)= \ha \cosh \chi\ \ln (\coth^2 {\chi\ov 2} ) - 1 \ .}
\eqn\adesso{ f_{ADS}(\chi) = 32  p a^4\
\bl( \ha \cosh \chi\ \ln (\coth^2 {\chi\ov 2} ) - 1\br) \ .}
Notice that in the Anti-De-Sitter case we did not have to `cut' the solution
by itroducing a $\Theta$-function because \adesso\ vanishes by itself for
large $\chi$. Of course this is due to the fact that in contrast to the
De-Sitter case the 2-dimensional space the shift function $f$ takes values on
is a non-compact hyperboloid and not the 2-sphere.
The corresponding `refraction function' is
\eqn\maria{ R_{ADS}(\chi)= 64 p a^4\
\bl( -{\cosh\chi \ov \sinh 2\chi} +
\ha \sinh\chi\ \ln (\coth^2 {\chi\ov 2})\br) \ .}
As a function $f_{ADS}(\chi)$ ($R_{ADS}(\chi)$) monotonically decreases
(increases) from plus (minus) infinity to zero as we get away from the pole
of the hyperboloid at $\chi =0$.
This is similar (but not exactly the same) to the behavior
of the corresponding functions in the previous section.

Let us now return to the more general case of the \SW-De-Sitter black hole
for which also there a positive cosmological constant.
The metric is
\eqn\sds{ ds^2_{SDS}=  - \bl( 1- {2M\ov r} - {r^2\ov a^2}\br)\ dt^2
+  \bl( 1- {2M\ov r} - {r^2\ov a^2}\br)^{-1} dr^2 +
r^2 (d\th^2 + \sin^2 \th\ d\phi^2)\ .}
Since it is a straightforward calculation we decided not to give
algebraic details concerning the derivation of the various results.
The differential equation to be satisfied by the shift function $f(\th)$
is again of the type (D.1) and it turns out that depending on the ratio $a/M$
there are two branches of solutions for the constants $c$ and $k$.
%corresponding to the
%different types of roots of a cubic algebraic equation.
We will only consider the branch in which the null surface where we will place
the massless particle corresponds to a positive value of $r$
(for positive $M$). In this branch
\eqn\csd{\eqalign{ &c = 2 \sin {\varphi\ov 3}\ (\sqrt{3} \cos {\varphi\ov 3}-
\sin {\varphi\ov 3} )\ ,\qq \cos \varphi \equiv \sqrt{27}\ {M\ov a} \cr
& {a\ov M} > \sqrt{27}\quad \Rightarrow \quad
 \varphi \in [-{\pi\ov 2}, 0) \cup (0,{\pi\ov 2}]\ ,\quad
c\in [-2,0)\cup (0,1]  \ ,    \cr }}
and the constant $k$ is always positive with precise value which
is not of any particular interest.
The null surface $u=0$ where the massless particle is placed corresponds to
$r=r_3=2{a\ov \sqrt{3}} \cos{\varphi +\pi\ov 3}$.
For positive (negative) values of
$\varphi$ the radial coordinate $r$ lies between
$ r_3 < r < r_1 $ ($r_1 < r < r_3$), where
$r_1=2{a\ov \sqrt{3}} \cos{\varphi -\pi\ov 3}$ ( $r_1$, $r_3$ and
$r_2 = -2{a\ov \sqrt{3}} \cos{\varphi \ov 3}$ are the three surfaces where
$g_{tt}$ vanishes).
The boundary cases $\varphi=-{\pi\ov 2},\ {\pi\ov 2}$ corresponding to the
De-Sitter space ($c=-2$) and the \SW\ black hole with zero cosmological
constant ($c=1$) have been already discussed.
The case $\varphi=0$ (${a\ov M}=\sqrt{27}$, $c=0$)
is excluded because the situation is similar
to the extremal \RN\ charged black hole.
%In the second branch
%\eqn\csdd{\eqalign{ & c=-1 -3 \bl({M\ov a}\br)^{2/3} (\L_+^2 + \L_-^2 )\ ,
%\qq \L_{\pm}= \bl( 1\pm \sqrt{1-{a^2\ov 27 M^2}}\br)^{1/3} \cr
%& {a\ov M} \leq \sqrt{27} \quad \Rightarrow \quad c\in (-\infty, -3]\ .\cr } }
Except for the case $c=-2$ the solution is given by (D.2). For the range
$0<c\leq 1$ the behavior of the solution is exactly the same as in the case
of the \RN\ charged black hole of the previous section.
For the range $-2<c<0$, similarly to the case of \fff, we find the
following integral representation
\eqn\fffi{ f(\th;c) = -{k\ov 2 c}
 -k \int_0^{\infty} ds \cosh(\sqrt{1/4-c}\ s)\ \bl( {1/\sqrt{2}\ov
 \sqrt{\cosh s - \cos \th }} - e^{-s/2} \br)\ .}
Comparing with \fff\ we see that
there is a second term in the integrand that essentially
regulates the divergent behavior of the first term for large values of $s$.
We note that the solution again blows up at $\th=0$ and as in the case of the
De-Sitter space it is monotonically increasing as we move from the northern
to the southern pole of the 2-sphere and changes from negative to positive
values at an angle $\th_0$ that depends on the value of $c$. For
instance for $c=-1$ we have $\th_0\simeq 61.59^0$.
Thus similarly to the De-Sitter
space case the magnitude of the shift function is reaching its minimum
at a point which is not the furthest from the massless particle at $\th=0$.
The important difference with
the De-Sitter space case is that the solution is extended to the southern
hemisphere as well, i.e. no use of $\Theta$-functions is required.
Let us also note that the `refraction function' is a monotonically decreasing
function of $\th$ and that the functional dependence in the region around the
northern and the southern poles is again given by \npole-\spolere\
(with a different value for $k$ and where
the appropriate analytic continuations
in the hyperbolic functions should be performed).

Let us also mention that we refrained from presenting here any
results concerning analogs of \RN\ charged black holes (see \KSHM) with
a cosmological constant because that would have complicated things rather
unnecessarily.

\subsec{ The 3-dimensional black hole }

It is interesting to consider the Anti-de-Sitter space in $d=3$.
Then the metric is
\eqn\itter{ ds^2 = -({r^2\ov a^2}-1)\ dt^2 + ({r^2\ov a^2}-1)^{-1} dr^2
+ r^2 d\chi^2 \ ,}
where $\chi$ is a non-compact coordinate. Again the constant $a$ is related
to the cosmological constant which is negative, i.e. $\L=-1/a^2$.
The change of variables that
brings \itter\ into the form \metrd\ is similar to \tutt\ with the lower
sign in the expression for $F(r)$. Then the
1-dimensional equation that $f$ satisfies using \condit\ for $d=3$
turns out to be
\eqn\deee{ {\partial^2 f\ov \partial \chi^2} \ - \ f \
= \ 64\pi p a^4 \ \d (\chi)\ .}
Its solution that has the correct
asymptotic behavior reads (we ignore the obvious solution of the
homogeneous equation)
\eqn\dsol{ f(\chi) = - 32\pi p a^4\ e^{-|\chi|} \ .}
A more physical situation arises when the non-compact variable $\chi$ is made
a compact one by letting $\chi\to \phi $ and identifying $\phi\equiv \phi + T$.
The most physical choice is of course $T=2\pi$.
In this case we have a black hole in $d=3$ as it was shown in \banados.
In fact this solution can also be thought of as a solution to string
theory with non-trivial antisymmetric tensor and
constant dilaton fields \hwka.
The corresponding exact conformal field theory
is obtained if one quotients
the $SL(2,\IR)$ WZW model by a discrete subgroup.
Because of the above discrete identification
the $\d$-function that appears in \deee\ should be replaced by
$\sum_{n=-\infty}^{\infty} \d (\phi - n T)$. Then instead of \dsol\ the
appropriate periodic solution satisfying $f(\phi + T)=f(\phi)$ is
\eqn\solpe{ f(\phi) = -32\pi p a^4
\sum_{n=-\infty}^{\infty} e^{-|\phi- n T|}\ ,\qq |\phi|< \infty \ .}
After some algebra the infinite sum is computed to give
\eqn\exxp{ f(\phi) = -64\pi p a^4\ \bl( {1\ov e^T -1}\ \cosh \phi  + \ha
 e^{-|\phi|} \br) \ ,\qq  |\phi| \leq {T\ov 2} \ .}
Notice that the range of $\phi$ is restricted in the fundamental domain and
that for $T\to \infty$ we recover the expression \dsol\ as we should.
On physical grounds we expect that the
absolute value of the shift is always larger that the one predicted
in the non-compact case, for any $\phi$. This should be the case because the
identification $\phi \equiv \phi + T$ `creates' many different sources
of infinite curvature ($\d$-functions) in the entire real line. Using
\exxp\deee\ one easily verifies that statement. In fact $|f(\phi)|$ reaches
its maximum (minimum) value at $\phi =0$ ($|\phi| = T/2$).
Finally, the `refraction function' is
\eqn\retd{ R(\phi)= -128\pi p a^4\ \bl( {1\ov e^T -1}\ \sinh \phi
- \ha \sign(\phi) e^{-|\phi|} \br) \ ,\qq  |\phi| \leq {T\ov 2} \ .}

%\newsec { Some string theory }
\newsec { String inspired gravitational solutions }

In low energy heterotic string theory we may consider the set of
background fields that comprises a metric, an antisymmetric tensor, a
dilaton and an
electromagnetic field coupled in a 2-dimensional $\s$-model action.
Then the requirement of conformal invariance generates constraints
these fields have to obey, the so called beta-function equations.
For our purposes it is convenient to present these equations in the
Einstein frame where the metric $g^E_{\m\n}$ is related to the $\s$-model
one by
\eqn\enm{ g^E_{\m\n}= \exp ({2\ov d-2} \Phi)\ g_{\m\n} \ .}
In the rest of this section we drop the extra index having in mind that all
quantities are evaluated in the Einstein frame.
To lowest order in the string coupling constant $\a'$, the
beta-function equations (see for instance \CALLAN) are\foot{ We will not
mention the dilaton beta function since it is not
independent from the other ones. Also since in the
examples we will shortly give the antisymmetric tensor is locally trivial,
i.e. $H_{\m\n\r}=0$, we restrict to only such cases although it is
straightforward to workout out the details in the more general case as well.}
\eqn\streinst{\eqalign{
&R_{\m\n} = {1\ov d-2}\bl(D_{\m}\Phi D_{\n}\Phi - g_{\m\n} D^2\Phi\br)
-{d\ov d-2}\bl(F_{\m\l}F_{\n}{}^{\l} -{1\ov d} g_{\m\n} F^2\br)
\exp({2\ov d-2}\Phi)  \cr
& D_{\l} \bl(\exp({2\Phi\ov d-2}) F_{\m}{}^{\l}\br) = 0\ .\cr }}
The first of the above equations is of the form of Einstein's
equations \Einst\ with non-trivial matter energy momentum tensor
that is being created by $\Phi$ and
$F_{\m\n}$ and given by the right hand side of this equation.
In the following we consider metric tensors of the form given by \metrd, and
electromagnetic field strength and dilaton field of the form
\eqn\fdil{\eqalign{
& F=2\ F_{uv}(u,v)\ du\wedge dv\ +\ F_{ij}(x)\ dx^i \wedge dx^j \cr
& \Phi= \Phi (u,v)\ .\cr }}
Strictly speaking with the above form for the electromagnetic field $F_{\m\n}$
the antisymmetric tensor
field strength $H_{\m\n\r}$ cannot be set to zero since
its source term $dH=-F \wedge F$ does not vanish. However, one
may think of having two independent electromagnetic fields
$F^a ,a=1,2$ given by
\eqn\ffdil{ F^1=2\ F^1_{uv}(u,v)\ du\wedge dv\ ,\qq
F^2= F^2_{ij}(x)\ dx^i \wedge dx^j \ .}
In that case $dH=-F^a \wedge F^a =0$. The two formulations are equivalent
since in both cases the electromagnetic energy momentum tensors are the same
leading to the same equations of motion given by \streinst.

As in section 1 we `perturb' a given background solution by adding a massless
particle moving along the $v$-direction at $u=0$.
Since the first equation in \streinst\ is the same as Einstein's equation
\Einst\ the conditions on the functions $g(u,v)$, $A(u,v)$ and $f(x)$ are
precisely
given by \condit, whereas $T_{vv}=0$ at $u=0$ translates into a condition
on the dilaton (again at $u=0$)
\eqn\addco{ \Phi_{,v} = 0 \ .}
Examination of the second equation in \streinst\ gives the condition
$(F_{uv})_{,v} = 0$ at $u=0$ which however is obeyed thanks to the equations of
motion and the forementioned conditions on $A(u,v)$ and $g(u,v)$.

\subsec{ An electrically and magnetically charged black hole}

As a first application of the above let us consider a
solution to low energy heterotic string theory
that represents a black
hole with both electric and magnetic charges \GiMa\KLOPV
\eqn\neem{ \eqalign{
&ds_E^2 = - \l_1 (r)\ dt^2 \ + \ \l_1^{-1} (r) \ dr^2 \
+ \ (r^2-r_0^2)\ (d\th^2 + \sin^2 \th\ d\phi^2)   \cr
& \Phi= \ln \bl({r-r_0 \ov r+ r_0}\br)  \cr
& F= {Q_E\ov (r-r_0)^2}\ dt \wedge dr \
+ \ Q_M \sin \th\ d\th \wedge d\phi\ ,\cr}}
where
\eqn\laena{\eqalign{ &\l_1 (r) = {(r-r_+)(r-r_-)\ov r^2 -r_0^2}\ ,\qq
r_0={Q_M^2 - Q_E^2\ov 2M} \cr
& r_{\pm} = M \pm \sqrt{M^2+r_0^2 -Q_E^2 -Q_M^2 }\ .\cr }}
As in the case of \rei, $r_+$ and $r_-$ are the outer and inner
horizons respectively. The constant $r_0$ is the so called dilaton charge
that essentially measures the difference between electric and magnetic
charges.
%For $r_0 =0$ (equal
%electric and magnetic charges) we recover the familiar charged
%\RN\ metric in $d=4$ (cf. \rei). For $r_-=r_0$ ($r_-=-r_0$) we have the
%case of zero electric (magnetic charge).
One easily finds that a transformation of the type \traa, but with
$\a = 2 (r_+^2 - r_0^2)/(r_+ - r_-) $, takes the metric into the form
\metrd\ with the relevant functions given by
\eqn\fagg{\eqalign{ &F(r)= e^{r/\a}
(r-r_-)^ { -{r_-^2 -r_0^2 \ov 2(r_+^2 - r_0^2)}} (r-r_+)^{\ha}
\ ,\qq A(u,v) = {\a^2\ov 2} \l_1 (r)/F(r)^2 \cr
& g(u,v)=r^2 - r_0^2\  . \cr } }
Obviously after this change of coordinates the expressions for the dilaton and
the electromagnetic field strength assume the form \fdil.
One easily finds that \condit\addco\ are satisfied at $u=0$ (or $r=r_+$)
and that the equation for the shift function $f$ is given by (D.1) with
\eqn\cooock{c= r_+ {r_+ - r_-\ov r_+^2 - r_0^2}\ ,\qq
k= 32  p (r_+^2-r_0^2)^2 (r_+ - r_- )^{-{r_+^2- r_-^2\ov r_+^2 - r_0^2}}
e^{-r_+{r_+ - r_- \ov r_+^2 -r_0^2}} \ .}
Similarly one easily sees that because of the fact that
$r_0 \leq r_- < r_+ $ we have $0<c \leq 1$ and therefore the solution to $f$
is given also by the general expression (D.2) or \fff.

%Let us also mention in pass that we can consider
%a larger class of `string inspired' four dimensional black holes that are
%similar to the one in this subsection but which have only electric charge
%and also contain a parameter a introduced via the term $e^{a\Phi} \Tr(F^2)$
%in the low energy string effective action.

%As a final remark let us mention that it is quite straightforward to
%work out explicitly the case of the $3d$ black string solution that is based
%on the coset conformal field theory $SL(2,\IR)\otimes \IR/\IR$ \HOHO. The
%result is that the shift function $f$ obeys an equation similar to \deee\
%but with the coefficient of the second term being $-2/(q+1)$ instead
%of $-1$ ($q$ is the embedding parameter) and some other constant $k_2$
%instead of $k_1$. Obviously its solution will be of the form \dsol\ or
%\exxp\ with the appropriate minor modifications.

\subsec{ The coset $SL(2,\IR)/\IR \otimes \IR^2 $ }

Finally let us consider the case of the 4-dimensional model that is the tensor
product of the 2-dimensional
coset $SL(2,\IR)_{-k}/\IR$ \WIT\ with 2 additional non-compact
dimensions. The metric in the Einstein frame (c.f. \enm) and the dilaton are
\eqn\tdbh{\eqalign{ &ds_E^2 = - 2\ \e\ du dv + (1-uv) (dx^2 + dy^2) \cr
& \Phi = \ln (1-uv) \ ,\cr }}
with $\e=\sign(k)$, where $(-k)$ is the central extension of the
$SL(2,\IR)_{-k}$ current algebra. For $\e =1$ the causal structure of the
spacetime is that of a black hole \WIT\
with a singularity at future times $t=u+v$, whereas for $\e=-1$ it has
the cosmological interpretation of an
expanding Universe with no singularity at future times $t=u-v$
(see \KOULU).
The metric is already of the form \metrd\ with
\eqn\tina{ A(u,v)= - \e \ ,\qq g(u,v)= 1-uv\ .}
It is easy to see that \addco\ is satisfied and that
the differential equation for the shift function is
%the
%modified Bessel one of zero order
%with the additional $\d$-function source term in the right hand side.
%Namely
\eqn\bessee{ ({d^2 \ov d\r^2} +{ 1\ov \r} {d\ov d\r} - \e ) f(\r)
= -16 \e  p\ {1\ov \r} \d(\r) \ ,}
where $\r^2=x^2 + y^2$.

Let us first consider the black hole case ($\e=1$). Then \bessee\ is nothing
but the modified Bessel equation of order zero
with the additional $\d$-function source term in the right hand side.
Its solution with the correct asymptotic behavior is
\eqn\solbee{\eqalign{& f_+(\r) = 16p\ K_0( \r) \cr
& R_+(\r)= 16 p\ K_1(\r)\ .\cr }}
In order to obtain a $\d$-function behavior in the right
hand side of \bessee\ one should carefully treat the derivative part
of the left hand side using the fact that $K_0(\r) \simeq -\ln (\r)$
for $\r<< 1$.
Also if one wishes to compactify one or both of the $x,y$ coordinates
one should replace the solution \solbee\ by a periodic one that will
necessarily involve infinite series as it was done in the case of
\dsol\ and \solpe.

In the cosmological case ($\e=-1$) \bessee\ becomes
the usual Bessel equation with solution respecting the singular behavior
of the source term in the right hand side given by
\eqn\sobee{\eqalign{& f_-(\r) = 8\pi p\ N_0( \r) \cr
& R_-(\r) = -8 \p p\ N_1(\r)\ ,\cr }}
where $N_n(\r)$ denotes Bessel functions of the second kind.\foot{ A more
systematic way to obtain the solution of \bessee\ is to expand
as $f(\r)=\int_0^{\infty} dk k A(k) J_0(k\r)$ and use the fact that
${\d(\r)\ov \r}= \int_0^{\infty} dk k J_0(k\r)$. Then easily we obtain
$A(k)={16 p \e \ov k^2+\e}$. Performing the integral we get for the shift
function the expression in \solbee\ or \sobee\ depending on the value of $\e$.}
Obviously \solbee\sobee\ are similar to \npole\npolere\ since in both cases
the metric in the transverse $x$-plane is the flat one
(notice:
$N_0(\r)\simeq {2\ov \pi} \ln(\r)$ and $N_1(\r) \simeq -{2\ov \pi} {1\ov \r}$
for $\r <<1$).
However, whereas in the
black hole case the shift and the `refraction function' never change sign,
in the cosmological case they do so infinite number of times until the
die off completely at $\r=\infty$. This type of behavior most likely
will show up in other cosmological models as well.
The difference in behavior is more dramatic
far away from the
particle's position at $\r=0$ since for $\r>>1$
$K_0(\r)\simeq \sqrt{{\pi \ov 2 \r}} e^{-\r}$ whereas
$N_0(\r)\simeq \sqrt{{\pi \ov 2 \r}} \sin(\r -{\pi\ov 4})$.

\newsec{Discussion and concluding remarks}

In this paper we have found the necessary and sufficient conditions for being
able to introduce a gravitational shock wave in a quite general class
of background solutions in Eistein's general relativity coupled to
non-trivial matter (with or without a cosmological constant)
and string theory. We have also applied the general
formalism to many cases that are of particular interest such as charged
black holes and constant curvature spacetimes.
The are also a few points we would like to further discuss.

1.
It is quite interesting to notice that the addition of the massless particle
in a background geometry creates in effect a perturbation that can be
described, in the context of string theory in curved spacetime,
in terms of a massless vertex operator. To be precise consider Einstein's
equations in the vacuum or in a bosonic string theory language a 2-dimensional
$\s$-model with zero antisymmetric tensor and dilaton fields. Let's take
\eqn\genop{V=F_{\m\n}(X)\del X^{\m} \bd X^{\n}\ , }
as a candidate for a massless vertex operator.\foot{ In general \cagan\ one
may add the term $\a' R^{(2)} F(X)$, where
$R^{(2)}$ is the curvature of the 2-dimensional worldsheet and $F(X)$ is a
tachyon-like operator. However this is not necessary here.}
If we want to perturb the $\s$-model by this operator we should require that
in every order in perturbation theory in the string coupling constant $\a'$,
its anomalous dimension is 2 so that the perturbation stays marginal.
That gives a set of consistency conditions $F_{\m\n}$ is constrained to
satisfy. To leading order in the string coupling they read \cagan
\eqn\cooo{\eqalign{ & -D^2F_{\m\n} - D_{\m}D_{\n} F_{\l}{}^{\l}
+ R_{\m}{}^{\r\s}{}_{\n} F_{\r\s} + D_{(\m} D^{\l} F_{\l\n)} =0 \cr
&  D^2 F_{\l}{}^{\l} - D^{\r}D^{\s}F_{\r\s} + R^{\r\s} F_{\r\s}=0\ .\cr}}
It is a straightforward (though a bit lengthly) calculation to verify that
\eqn\verrr{ F_{uu}= -2 A(u,v) f(x) \d(u) \ ,}
as it is read from \metrhat,
is indeed a massless vertex operator corresponding to the background metric
\metrd\ provided that, as in \condit, at $u=0$ we have $\gpv=0$ and
$\Apv=0$ and that $f(x)$ satisfies the {\it homogeneous}
differential equation in \condit. The vanishing of the right hand side
occurs because we have not included the energy momentum tensor
corresponding to this vertex \enepart. This is not usually done if the
background is flat because then the resulting homogeneous equation has a
solution (with a different anzatz for a vertex operator (see \FMS)),
which is not however always true, for instance, as we have seen
in the cases of black holes.

2.
What is the meaning of the diverging solution (see \fff)
as we go to the extremal limit of a \RN\ charged black hole and moreover the
no solution result we have found working exactly at the extremal limit?
In general the existence of a solution means that the original background
responds to the `perturbation' created by the massless particle.
A particle geodesic, in a charged close to the
extreme limit black hole, gets shifted
and refracted by large amounts due to the large gravitational field created
by the massless particle. Exactly at the extremal limit there is no solution
with a {\it single} particle but there is one if in
addition to the {\it positive} energy massless particle at the northern pole
we place another one with {\it negative} energy of the same magnitude at the
southern pole (both at $r=r_+=r_-$).
Then the solution with the correct singular behavior at $\th=0,\pi$ is
$f(\th)=-k Q_0(x)=-{k\ov 2} \ln\bl(\cot ^2{\th\ov 2}\br)$.
In other words we might think that as $r_+\to r_-$ and the gravitational field
becomes infinite it is necessary to take the {\it antiparticle} into
consideration as well.
The fact that there is a solution in the case of an extremal \RN\ black hole
only if we use a {\it particle} and its {\it antiparticle} is somewhat
analogous to the `semiclassical' explanation of Kleins's paradox
using the concept of a `sea' filled with negative energy particles.

As a final comment on the \RN\ charged black hole let us mention what the
result would be if we had placed the massless particle in the inner horizon
instead of the outer one. The differential equation for the shift function
is again of the form (D.1) with
\eqn\coocck{c=1-{r_+\ov r_-}\ \ ( -\infty < c < 0 )\ ,\qq
k= -32  p r_-^4 (r_+ - r_- )^{{r_+^2- r_-^2\ov r_-^2}}
e^{{r_+ - r_- \ov r_-}} \ .}
Thus $c$, $k$ are basically given by \coock\ with $r_+$ and $r_-$
interchanged. The solution for the shift function is given by (D.2) which one
can show is equivalent to
\eqn\magic{\eqalign{ &f_n(\th) = -k \sum_{l=0}^{n-1} {l+\ha \ov l(l+1) + c}\
P_l(\cos \th) - k \int_0^{\infty} ds\ \cosh(\sqrt{1/4-c}\ s)\
\bl({1/\sqrt{2}\ov \sqrt{\cosh s - \cos \th}}\cr
& - \sum_{l=0}^{n-1} e^{-(l+\ha) s}\
P_l(\cos\th) \br)\ , \qq  -n(n+1) < c < -n(n-1), \ n=1,2\dots \ .\cr }}
Using \magic\ one finds that the shift function
starts from plus infinity at $\th=0$ becomes
zero and changes sign as many times as the value of $n$ and eventually assumes
the finite constant value predicted by \spole.
Notice also that \fffi\ corresponds to the case $n=1$ of the above formula.
Of course for the cases where $c=-n(n+1),\ n=1,2,\dots $
(D.5) or (D.6) gives the
solution. The case $c=-2$ ($n=1$) is exactly the same as in the De-Sitter
space and the solution is unique. For higher values of $n$ however, there
are ($n-1$) undetermined coefficients which can probably be fixed with some
additional physical requirements for the shift or the `refraction function'.

3. Can we find the shock wave-type of disturbances created by two or more
massless particles moving along parallel surfaces
(different values of $u$) just by a coordinate shift?
Obviously this requires $\Apv=\gpv =0$ at these
values of $u$. This is not possible for black hole
type solutions. It can however be done for backgrounds with a covariantly
constant null Killing vector (see Appendix B) that include plane waves and the
flat space as special cases. Then at each null surface we have a shock wave
with a shift function obtained by solving (B.6).

Finally, it would be obviously very interesting to explicitly compute,
using techniques similar to those in \hoo\VV\AGV, particle
scattering amplitudes in the shock wave curved geometries we have derived.
In particular one would like to know how these amplitudes depend on the
different behaviors of the shift and the `refraction function' we have
demonstrated in the present paper.

\vfill\eject

\centerline{ \bf Acknowledgments}

I would like to thank I. Bakas, G. 't Hooft and E. Verlinde
for useful discussions, B. de Wit for reading the manuscript and C. Loust\'o
for e-mail correspondence.
I would like also to thank CERN for its hospitality during which part of
this work was done and the anonymous referee for verifying the expression for
the Ricci tensor (A.2).

\bs\bs

\centerline { \bf Note added }

Towards finishing typing this paper we become aware that the authors of
\lathos\
dealt with some of the issues of this paper. However, we disagree with many of
their findings. The main points of disagreement are
(equation labels starting with `LS.' refer to \lathos):

1) For the general $d$-dimensional case we do not agree with the expressions
for the Ricci tensor (LS.5-9). Our (A.2)
contains some additional terms with explicit $d$ dependence.

2) We do not quite agree with the conditions (LS.17-18) (we find instead
\condit) mainly because of the explicit
appearance of the cosmological constant in (LS.18) and the absence of a
condition on $T_{vv}$ in (LS.17).

3) For the case of a \RN\ charged black hole (LS.35) agrees with (D.2)
(with $a=c$). However, we find the different integral representation \fff\
instead of (LS.36).

\vfill\eject

\appendix A  {Components of useful tensors  }

The corresponding to the metric \metrhat\ non-vanishing
\CS\ symbols are (for notational convenience we omit the hats)
\eqn\crist{\eqalign{ & \G^u_{uu}= - {\Fpv\ov 2A}+ {\Apu\ov A} \ ,\qq
\G^u_{ij}= -{\gpv\ov 2A}\ h_{ij} \cr
& \G^v_{uu}= {\Fpu\ov 2A} + {F \Fpv \ov 2 A^2} - {F \Apu\ov A^2}\ ,\qq
\G^v_{uv} ={\Fpv \ov 2A} \ ,\qq \G^v_{ui} = {\Fpi\ov 2 A}\cr
& \G^v_{vv}= {\Apv\ov A} \ ,\qq \G^v_{ij}= \bl(-{\gpu \ov 2 A}
+{F \gpv\ov 2 A^2}\br)\ h_{ij} \cr
&\G^i_{uu}= -{1\ov 2 g}\ h^{ik} F_{,k}\ ,\qq \G^i_{uj}= {\gpu\ov 2 g}\ \d^i_j
\ , \qq \G^i_{vj}= {\gpv\ov 2 g}\ \d^i_j \cr
& \G^i_{jk}= \ha h^{il}( h_{lk,j} + h_{lj,k} - h_{jk,l}) \ .\cr }}
Using the above expressions we find that the non-vanishing components
of the Ricci tensor are (we substitute $F=-2 A f \d $ (see \metrhat))
\eqn\ricci{\eqalign{ &R_{uu}= {d-2\ov 2}\ \bl( {\gpu \Apu\ov g A} -
{g_{,uu} \ov g} + {\gpu^2 \ov 2 g^2} \br)
\ +\ {A\ov g}\ \d\ \triangle_{h_{ij}} f\ -\ {d-2\ov 2}\ {\gpv \ov g}\ \d' \ f
\cr
& \qq
 +\ \bl(2\ {A_{,uv}\ov A} - 2\ {\Apu \Apv\ov A^2} + {d-2\ov 2 g A}\ (\gpu \Apv
+ \gpv \Apu) \br)\ \d\ f  \cr
& \qq
 +\ 2\ \bl({A_{,vv}\ov A}- {\Apv^2\ov A^2} + {d-2\ov 2}\ { \gpv\Apv \ov g
A}\br)\
 \d^2\ f^2 \cr
&R_{uv}= \bl( {\Apu \Apv\ov A^2} - {A_{,uv}\ov A}
+ {d-2\ov 4}\ {\gpu\gpv\ov g^2} - {d-2\ov 2}\ {g_{,uv}\ov  g } \br) \cr
&\qq
+\ \bl({\Apv^2\ov A^2}- {A_{,vv}\ov A} -{d-2\ov 2}\ {\gpv\Apv\ov gA}\br)\
 \d\ f\cr
& R_{ui}= -\bl( {d-4\ov 2}\ {\gpv\ov g} + {\Apv\ov A} \br)\ \d\ \fpi  \cr
&R_{vv}= {d-2\ov 2}\ \bl({\gpv \Apv\ov gA} + {\gpv^2\ov 2 g^2}
- {g_{,vv}\ov g} \br) \cr
&R_{ij}= R^{(d-2)}_{ij}\ -\
\bl( {d-4\ov 2}\ {\gpu\gpv\ov gA}+ {g_{,uv}\ov A}\br)
\ h_{ij}\ -\ \bl({d-4\ov 2}\ {\gpv^2\ov gA} + {g_{,vv}\ov A}\br)\ h_{ij}\
 \d\ f \ .\cr }}
In the 4-dimensional case \ricci\ reduce to the Ricci tensor given in \DHooft.
In the case we are dealing with string theory as in section 5 it might be
useful to compute the tensor $D_{\m}D_{\n} \Phi$ with $\Phi=\Phi(u,v)$.
Its non-vanishing components read
\eqn\comp{\eqalign{
& D_u D_u \Phi = ( \Phi_{,uu}- {\Apu \ov A} \Phi_{,u})\ - \
{1\ov A}\ ( \Apu \Phi_{,v} + \Apv \Phi_{,u})\ \d \ f \ + \ \Phi_{,v}\ \d'\ f
\ - \ 2 {\Apv \Phi_{,v} \ov A}\ \d^2 \ f^2 \cr
& D_u D_v \Phi = \Phi_{,uv} + {\Apv \Phi_{,v} \ov A}\ \d \ f \ , \qq
D_u D_i \Phi = \Phi_{,v}\ \fpi\ \d \cr
& D_v D_v \Phi = \Phi_{,vv} - {\Apv \Phi_{,v}\ov A} \cr
& D_i D_j \Phi = {1\ov 2 A}\ (\gpu \Phi_{,v} + \gpv \Phi_{,u})\ h_{ij}
\ + \ {\gpv \Phi_{,v} \ov A}\ h_{ij} \ \d\ f \ .\cr }}

\appendix B { Backgrounds with a covariantly constant null Killing vector }

Consider the metric with a covariantly constant null Killing vector
\eqn\metcov{ ds^2 = 2 dudv \ +\ g_{ij}(u,x)\ dx^i dx^j\ ,}
where $(i,j=1,2,\dots, d-2)$. The most general matter field energy momentum
tensor consistent with the non-vanishing components of the Ricci tensor
corresponding to \metcov\ (see (B.5) below) has the form
\eqn\enplane{T= T_{uu}(u,x)\ du^2 + 2\ T_{ui}(u,x)\ dudx^i +
 T_{ij}(u,x)\ dx^i dx^j \ .}
The analog of \metrhat\ is
\eqn\metrhht{\eqalign{& ds^2 = 2\ d\hat u d\hat v\ + \
\hat F\ d\hat u^2\ +\
 \hat g_{ij}\ d\hat x^i d\hat x^j\ , \cr
& \hat F= F(\hat u, \hat x)= -2\ \hat f\ \hat \d\ ,\cr } }
whereas that of \enhat\ is exactly the same tensor \enplane\
since no $v$-component
of the energy momentum tensor appears in \enplane.

The non-zero components of the \CS\ symbols and the Ricci tensor are
(once again we drop the hats and also we denote derivatives with respect to
$u$ with a dot)
\eqn\cripl{\eqalign{& \G^v_{uu}= \ha\ \dot F\ ,\qq \G^v_{ui}= \ha\ \Fpi\ ,
\qq \G^v_{ij}= -\ha\ \dot g_{ij} \cr
& \G^i_{uu}= - \ha\ g^{ik} F_{,k}\ ,\qq   \G^i_{uj}= \ha g^{ik} \dot g_{jk}\ ,
\qq \G^i_{jk}=\ha g^{il} (g_{lk,j} + g_{lj,k} - g_{jk,l})\ ,\cr }}
and
\eqn\riccipl{\eqalign{ & R_{uu}= \bl(-\ha \ddot g_{ij} g^{ij} + {1\ov 4}
g^{ik} g^{jl} \dot g_{ij} \dot g_{kl}\br) \ +\
\d\ \triangle_{g_{ij}} f\cr
& R_{ui}= - D_{[i} \dot g_{k]j} g^{jk}          \cr
& R_{ij}= R^{(d-2)}_{ij} \ .\cr } }

Then the condition to be satisfied at $u=0$ is the differential equation
(again we drop the hats over the symbols)
\eqn\cont{\triangle_{g_{ij}} f= 32\pi \ p\ \d^{(d-2)}(x)\ ,\qq
\triangle_{g_{ij}} = {1\ov \sqrt{g} }\del_i g^{ij} \sqrt{g} \del_j \ .}

Let us next consider string backgrounds with a covariantly
constant null Killing vector, i.e. with metric in the Einstein frame
as in \metcov, and in addition with
non-trivial antisymmetric tensor field strength components
$H_{ijk}=H_{ijk}(u,x)$, $H_{iju}=H_{iju}(u,x)$ as well as a dilaton
field $\Phi=\Phi (u,x)$. It can easily be shown that the `effective' matter
energy momentum tensor due to these background fields is again of the
form \enplane. Thus, the equation for the shift function is again given
by \cont, where the metric is the Einstein frame one.
For completeness, the non-vanishing components of $D_{\m}D_{\n}\Phi$ are
\eqn\tenpl{\eqalign{&D_u D_u \Phi  = \ddot \Phi \ - \
\d\ g^{kl} \Phi_{,k} f_{,l}  \cr
& D_u D_i \Phi =  \dot \Phi_{,i} - \ha g^{kl} \dot g_{li}  \Phi_{,k}
\cr
& D_i D_j \Phi = \Phi_{,ij} - \G^k_{ij} \Phi_{,k} \ .\cr }}
%it is very easy to show that the equation for
%$f$ is again given by \cont\ but with a different Laplacian that
%contains the dilaton in its definition,
%i.e. $\tilde \triangle_{g_{ij}} = {1\ov \sqrt{g} e^{\Phi}}
%\del_i \sqrt{g} e^{\Phi} g^{ij} \del_j $.
Let us also note that for the particular case of plane waves where all
scalars and tensors depend only on the light cone coordinate $u$ and not
on the $x^i$'s, equation \cont\ is the same as the one corresponding to the
flat space case and therefore its general solution is \aisexl\lousan
\eqn\flatt{ f(\r)=\cases{ 16  p\ \ln \r\ , \quad &if $d=4$
\cr \noalign{\vskip 6pt}
-{32 \pi  p \ov (d-4) \Om_{d-2}}\ {1\ov \r^{d-4}}\ ,  \quad &if $d>4$\ ,\cr }}
where in general $\Om_D={2\pi^{D/2}\ov \G(D/2)}$ is the `area' of the unit
sphere in $D$-dimensions and $\r^2\equiv g_{ij}(0)x^i x^j$.
Since we can freely shift $u$ by a constant the choice of $u=0$
should be made in such a way that the metric is non-degenerate at that point,
i.e. $u=0$ is not a focusing point of null rays.

\appendix C { Geodesic equations }

For the metric \metrhat\ the geodesic equations obtained by varying $v$
and $x^i$ are (in this Appendix the dots over the various symbols denote
derivatives with respect to the affine parameter $\tau$)
\eqn\geodess{\eqalign{ & \ddot u \ +\  {\Apu \ov A}\ \dot u^2 \ -\
{\gpv \ov 2A}\
h_{ij} \dot x^i \dot x^j \   +\  f\ {\Apv \ov A}\ \d\ \dot u^2 \  = \ 0 \cr
& \ddot x^i \ + \ \G^i_{jk} \dot x^j \dot x^k \ + \ {\gpu\ov g}\
\dot u \dot x^i \
+ \ {\gpv \ov g}\ \dot v\dot x^i \   + \
{A\ov g}\ \d\ f_{,j} h^{ji}\ \dot u^2 = 0 \ ,\cr }}
whereas the one obtained from the variation of $u$ is
\eqn\geoo{\eqalign{ & \ddot v \ +\ {\Apv \ov A}\ \dot v^2 \ -  \
{\gpu \ov 2 A}\ h_{ij} \dot x^i \dot x^j \cr
& + \bl( f {\Apu\ov A}\ \dot u^2 \ -\ 2 f {\Apv\ov A}\ \dot u\dot v \ - \
2 f_{,i}\ \dot u \dot x^i \ - \ {\gpv\ov A}\ f\ h_{ij} \dot x^i \dot x^j \br)
\ \d \cr
& -f\ \d'\ \dot u^2   \ + \ 2 f^2\ \d^2\  {\Apv\ov A}\ \dot u^2 \ = \ 0 \ .
\cr } }
{}From \metrhat\ the energy corresponding to the geodesic is given by
\eqn\enee{ 2A \dot u \dot v \ + \ g h_{ij} \dot x^i \dot x^j \ - \
2 f \d A \dot u^2 \ = \ \a \ ,}
where $\a = -1,0,1$ depending on whether the geodesic is timelike, null or
spacelike respectively. It is clear from \enee\ that the $\dot u^2$-term
can be compensated by a discontinuity in $v$ at $u=0$.
Taking $v=v_0 + \Theta(u) \Delta v$, where $v_0$ is a solution of the geodesic
equations for $f=0$, and integrating \enee\ over a small interval around
$u=0$ we obtain immediately the relation \shifftt. Performing the same
integration in the first of \geodess\ gives no new information because $\Apv=0$
at $u=0$. However the same procedure in the second equation in \geodess\ gives
\eqn\reee{ {\dot x^i}\big |_{u=0^-}  - {\dot x^i}\big |_{u=0^+}  =
{A \ov g}\big |_{u=0}\ f_{,j} h^{ji}\dot u \ .}
Our next step is to express the variation with respect to the affine
parameter $\tau$ at $u=0$ as a variation with respect to $u$ itself.
This can be done by noticing that the first equation in \geodess\ evaluated at
$u=0$ is
\eqn\geooo{ \ddot u + {\Apu\ov A} \dot u^2 \big |_{u=0} = 0\ ,}
and has as its solution
\eqn\ssoo{\dot u = {1\ov A} \big |_{u=0} \ .}
Therefore \reee\ takes the form of \refra. It is important to emphasize that
\ssoo\ is really a solution of \geooo\ only at $u=0$ where $\Apv=0$. Finally,
integrating \geoo\ over a small interval around $u=0$ gives no additional
conditions as it eventually reduces to \geooo.

\appendix D { Solutions of the equation for the shift function f }

Let us consider the differential equation
\eqn\eqfd{\eqalign{ &
 \triangle_{(2)} f \ - \ c\ f\ = \ 2\pi k\ \d(x-1)\d (\phi) \cr
& \triangle_{(2)}= \partial_x (1-x^2) \partial_x + {\partial_{\phi}^2
\ov 1-x^2}\ ,\qq x=\cos \th \ ,\cr }}
on the unit 2-sphere with metric
$ds^2 = d\th^2 + \sin ^2 \th\ d\phi^2$, where $k$, $c$ are real constants.
Except for the $\d$-function in the right hand side this is nothing but the
usual Legendre equation of order $\n$, where $\n$ is a solution of
$\n(\n+1)+c=0$.
Therefore its solutions depend very much on the value of the constant $c$.

We first consider the case of $c\neq -n(n+1)$, $n=0,1, \dots $.
Since the eigenvalues of the Laplacian operator are $E_n=-n(n+1)$,
$n=0,1,\dots $ the
modified Laplacian $(\triangle_{(2)}-c)$ has no zero modes and
therefore it is invertible. The general solution of \eqfd\ can be
given in terms of Legendre polynomials as
\eqn\soluu{ f(\th;c) =
- k \sum_{l=0}^{\infty} {l+\ha\ov l(l+1) + c}\ P_l(\cos \th)\ ,\qq
c\in \IR-\{-n(n+1)\ ,n=0,1,\dots \}\ .}
Since in this case $c$ does not coincide with an eigenvalue of the
Laplacian there is no solution to the homogeneous equation in \eqfd.

If $c=-l(l+1)$, $l=0,1, \dots$,
then we should make use of the eigenfunctions of the
Laplacian that are not regular at the poles at $x=\pm 1$,
namely $Q_l(x)$ that are defined as
\eqn\qlll{ Q_l(x)= \ha P_l(x) \ln \bl({1+x\ov 1-x}\br)
- \sum_{m=1}^l {1\ov m} P_{m-1}(x) P_{l-m}(x)\ .}
If the solution for $f$ is proportional to such a function then a
careful treatment of the derivative part in \eqfd\ will produce
$\d$-functions at the poles of the sphere at $x=\pm 1$. However according
to the right hand side of \eqfd\ there
should not be any singularity at $x=-1$. Therefore we are forced to
consider a solution of the form $f\sim Q_l(x)\ \Theta(x-x_0)$, where
$x_0$ is to be determined by requiring that $f$ satisfies \eqfd. After
some algebraic manipulations and using the fact that ${d\ov dx}
\Theta(x) = \d (x)$ we obtain the condition
\eqn\dete{ (1-x^2) {d Q_l(x)\ov dx}\ \d (x-x_0) = 0\ .}
Therefore $x_0$ must be one of the points where the slope of $Q_l(x)$
vanishes. In fact there are $l$ such points $\{x_1,x_2, \dots x_l\}$,
symmetrically distributed around $x=0$ which always satisfies \dete\
if $l$ is odd.
Thus the general solution is even by
\eqn\soql{\eqalign{ &f(\th)= - k\ Q_l(x)\ \bl( A_0 \Theta(x)
+ \sum_{m=1}^{{l-1\ov 2}} [ A_m
\Theta(x-x_m) + B_m \Theta(x+x_m)  ] \br) \ , \qq x=\cos \th \cr
& A_0+ \sum_{m=1}^{{l-1\ov 2}} (A_m + B_m) =1 \ ,\qq  l=1, 3, \dots \cr}}
if $l$ is odd and by
\eqn\seql{\eqalign{ &f(\th)= - k\ Q_l(x)\  \sum_{m=1}^{{l\ov 2}} \bl(A_m
\Theta(x-x_m) + B_m \Theta(x+x_m) \br) \ , \qq x=\cos \th \cr
& \sum_{m=1}^{{l\ov 2}} (A_m + B_m) =1 \ , \qq l=2, 4, \dots \cr }}
if $l$ is even. The $A_m$'s and the $B_m$'s are arbitrary constants.
The constraints on their sums to be unity is necessary because at
$x\simeq 1$ the
solution should behave like $f=-k Q_l (x) \simeq {k\ov 2} \ln (1-x) $ in
order to have the correct normalization. Notice that the solutions
\soql\seql\ are discontinuous at the points $x_m$ and
that for $l=0$ there is no solution (unless $k=0$) because there
exist no point where the slope of $Q_0(x)$ is zero.
It should be noted that we can add in \soql\seql\ any solution of the
homogeneous equation in \eqfd, namely a linear combination of the spherical
harmonics $Y^l_m(\th,\phi)$, $m=-l,-l+1,\dots ,l$.

Let us next consider the differential equation \eqfd\ but with
Laplacian
\eqn\lapl{ \triangle_{(2)}= \partial_x (x^2-1) \partial_x + {\partial_{\phi}^2
\ov x^2-1}\ ,\qq x=\cosh \chi \ , }
on the hyperboloid with the Lobachevsky metric
$ds^2 = d\chi^2 + \sinh ^2 \chi\ d\phi^2$. We consider the case when
$c=l(l+1)$, $l=0,1, \dots $. Then the solution is given by
\eqn\lysis{ f(\chi) = k\ Q_l(x) \ ,\qq x=\cosh \chi \ , \qq
l=0,1,\dots\ .}
Notice that in this case there is a solution for $l=0$.
Also we did not need to use any $\Theta$-functions because $f(\chi)$
as given by \lysis\ becomes asymptotically zero for large $\chi$ as it
should be.

Finally let us present the solution of the equation for the shift function
\condd\ which is valid in any number of dimensions $d \geq 4$. This can
be written down using the generalized spherical functions \BAIT, which are
representations functions of $SO(d-1)$. If the
number of spacetime dimensions is even they are defined in terms of usual
Legendre polynomials as
\eqn\gene{ P^{(2r)}_l(\cos \th) = {l+r-\ha \ov (2\pi)^r}\ \bl({d\ov
d \cos\th}\br)^{r-1}\ P_{l+r-1}(\cos \th)\ , \qq d=2r+2 ,\ r=1,2\dots \ ,}
whereas if it is odd in terms of Chebishev polynomials as
\eqn\geno{ P^{(2r-1)}_l(\cos \th) = 2 {l+r-1 \ov (2\pi)^r}\ \bl({d\ov
d \cos\th}\br)^{r-2}\ {\sin (l+r-1)\th \ov \sin \th}\ ,
\qq d=2r+1,\ r=2,3\dots \ .}
The only properties of the generalized spherical harmonics we need here are
\eqn\propp{\eqalign{
&\triangle_{(d-2)} P^{(d-2)}_l = -l(l+d-3) P^{(d-2)}_l \cr
& \d^{(d-2)}(x) =\sum_{l=0}^\infty P^{(d-2)}_l(\cos \th)\ .\cr }}
Using the above formulae we find that the solution for the shift function is
\eqn\soooll{ f_d(\th;a) = - 2 \pi b \sum_{l=0}^\infty {P^{(d-2)}_l(\cos \th)\ov
l(l+d-3) + a}\ .}
Obviously the above solution is not valid if $a$ coincides with one of the
eigenvalues of the Laplacian, i.e. if $a= -n(n+d-3), \ n=0,1\dots $.
Then one has to use the analogs of the $Q_l$'s (see \BAIT) but we will not
elaborate on that any more here.

\vfill
\listrefs
\end

\centerline { \bf Note added }

Towards finishing typing this paper we become aware that the authors of
\lathos\
dealt with some of the issues of this paper. However, we disagree with many of
their findings.
We will explain the main points of disagreement. In the following, equation
labels starting with `LS.' refer to \lathos.

1). The expressions for the Ricci tensor (LS.5-9) are just the expressions of
\DHooft, which however are not valid if $d\neq 4$.
We find instead (A.2).

2) We do not quite agree with the conditions (LS.17-18).
Firstly because of the explicit
appearance of the cosmological constant in (LS.18) and we find instead \condit.
A simple argument why
there should not be a cosmological constant dependent term in the equation
for the shift function is the following:
The effects of a non-vanishing cosmological constant can be reproduced by
a {\it matter} energy momentum tensor that is proportional to the metric
(see section 4). But if in (LS.18) we set $\L=0$ the equation can not
be reproduced by such a matter energy momentum tensor because it does not
depend at all on that. Also the particular result
of \HOTA\ is in agreement with \condit\ and not with (LS.18).
Secondly in (LS.17) there is no
condition on $T_{vv}$ and for $d\neq 4$ the wrong Ricci tensor was used.
%(LS.17-18) agrees with the correct one \condit\ only in $d=4$,
%for $\Lambda =0$ and if the entire
%dependence on $u,v$ is in the form $uv$.

In \lathos\ (as in this paper) the Einstein equations were used in
the coordinate system (SL.4) (or \metrhat). That means that
the components of the matter energy
momentum tensor although initially given in the system (SL.2) (or \metrd)
should be transformed in  that coordinate system (SL.4) (or \metrhat).
This was not done in \lathos\ (we did that in \enhat). Therefore certain terms
containing $\d$-functions were missed.
This we believe was the main slip in \lathos.

3) For the case of a \RN\ charged black hole (LS.35) agrees with (D.2)
(with $a=c$).
However, we disagree with
the integral representation (LS.36) which apparently was used to
draw fig.1 of \lathos.
% is incorrect (this mistake has been propagated in
%the literature (see eq. (37) of \lou).
Notice that although (LS.35) diverges (correctly) if
$a=0\ \forall\ \th$, (LS.36) does not.
%It is correct only for $a=1$ (\SW\ case \DHooft).
Also (LS.35) does not distinguish between $a\leq 1/4$ and $a> 1/4$.
We find instead \fff.

4) For the case of the De-Sitter space the authors of \lathos\
stated that there
is no solution for the shift function $f(x)$.
However, we have found the solution (see section 3 and paragraph
just before \adesso). Also we disagree with the result of \lathos\
for the case of the \SW-De-Sitter black hole.

\newsec { Some string theory }

The analog of Einstein's equations \Einst\ for heterotic string theory
in the presence of
non-trivial metric, antisymmetric tensor, dilaton and
electromagnetic field are given by the lowest order,
in the string coupling constant $\a'$, beta-function equations
(see for instance \CALLAN)
(we will not mention the dilaton beta function since it is not
independent from the other ones)
\eqn\streinst{\eqalign{
&R_{\m\n} - D_{\m} D_{\n} \Phi  - 2 F_{\m\l} F_{\n}{}^{\l}
- {1\ov 4} H_{\m\r\l} H_{\n}{}^{\r\l} =0 \cr
& D_{\l} (e^{\Phi} F_{\m}{}^{\l}) + {1\ov 12} e^{\Phi} H_{\m\r\l}
F^{\r\l} = 0 \cr
& D_{\l} (e^{\Phi} H^{\l}{}_{\m\n})=0 \ .\cr }}
One may think of the first equation in \streinst\ as the
generalization of the Einstein's equations \Einst\ in the presence of
non-trivial matter energy momentum tensor
that is being created by $\Phi$, $H_{\m\n\r}$ and
$F_{\m\n}$. The comparison is being made in the Einstein frame
where the metric $G^E_{\m\n}$ is related to the $\s$-model one by
\eqn\enm{ G^E_{\m\n}= \exp ({2\ov d-2} \Phi)\ G_{\m\n} \ .}
and where indeed the first equation in \streinst\ takes the
form \Einst\ (see \CALLAN).
In the following we consider metric tensors of the form given by \metrd, and
electromagnetic field strength and dilaton field of the form
\eqn\fdil{\eqalign{
& F=2\ F_{uv}(u,v)\ du\wedge dv\ +\ F_{ij}(x)\ dx^i \wedge dx^j \cr
& \Phi= \Phi (u,v)\ .\cr }}
Since in the examples we will shortly give the antisymmetric tensor is trivial,
i.e. $H_{\m\n\r}=0$, we restrict ourselves to only such cases although it is
straightforward to workout out the details in the more general case.

As in section 1 we `perturb' a given background solution by adding a massless
particle moving along the $v$-direction at $u=0$.
In principle one has to repeat the computation that lead to the
condition \condit\ but now for the set of equations \streinst.
However, the conditions on the metric components can be obtained by
going in the Einstein frame and in fact are given exactly by \condit\
with the functions $g(u,v)$ and $A(u,v)$ replaced by the corresponding
Einstein frame expressions
\eqn\gaein{ A^E(u,v)= \exp ({2\ov d-2} \Phi)\ A(u,v)\ ,\qq
g^E(u,v)= \exp ({2\ov d-2} \Phi)\ g(u,v)\ .}
However, there is an additional constraint obtained by examing the
second equation in \streinst\ (with zero antisymmetric tensor field
strength) for $\m = u$. One easily finds that
\eqn\addco{ \bl({F_{uv}\ov A^E} \br)_{,v} = 0 \ ,}
should also be satisfied at $u=0$.